\documentclass[aps,pre,twocolumn,nofootinbib,floatfix,showpacs]{revtex4}  %PRE format

\usepackage{color}
\usepackage{soul}

\usepackage{graphicx} %Include figure files
\usepackage{amssymb}
\newcommand {\be}{\begin{equation}}
\newcommand {\ee}{\end{equation}}
\newcommand {\ba}{\begin{eqnarray}}
\newcommand {\ea}{\end{eqnarray}}

\newcommand{\Jm}{J_{\mathrm{m}}}
\newcommand{\Om}{\omega_{\mathrm{m}}}

\newcommand{\Oml}{\omega_{\mathrm{m}}^{_{(L)}}}
\newcommand{\Omr}{\omega_{\mathrm{m}}^{_{(R)}}}
\newcommand{\kl}{k_{_L}}
\newcommand{\kr}{k_{_R}}
\newcommand{\kc}{k_{_C}}
\newcommand{\ko}{k_{_0}}
\newcommand{\Vl}{V_{_L}}
\newcommand{\Vr}{V_{_R}}
\newcommand{\bl}{b_{_L}}
\newcommand{\br}{b_{_R}}
\newcommand{\nc}{n_{_c}}
\newcommand{\Jl}{J_{_L}}
\newcommand{\Jr}{J_{_R}}

\begin{document}

%\draft
%\wideabs{

\title{Effect of external potential on the energy transport in harmonically driven segmented Frenkel-Kontorova lattices}

\author{M.~Romero-Bastida}
\affiliation{SEPI ESIME-Culhuac\'an, Instituto Polit\'ecnico Nacional, Av. Santa Ana No. 1000, Col. San Francisco Culhuac\'an, Culhuac\'an CTM V, Coyoacan, CDMX 04440, Mexico}
\email{mromerob@ipn.mx}
%\email{}

\date{\today}

\begin{abstract}
Thermal resonance, that is, the heat flux obtained by means of a periodic external driving, offers the possibility of controlling heat flux in nanoscale devices suitable for power generation, cooling, and thermoelectrics among others. In this work we study the effect of the onsite potential period on the thermal resonance phenomenon present in a one-dimensional system composed of two dissimilar Frenkel-Kontorova lattices connected by a time-modulated coupling and in contact with two heat reservoirs operating at different temperature by means of molecular dynamics simulations. When the periods of the onsite potential on both sides of the system are equal the maximum resonance is obtained for the lowest considered value of the period. For highly structurally asymmetric lattices the heat flux toward the cold reservoir is maximized, and asymmetric periods of the onsite potential afford an extra way to control the magnitude of the heat fluxes in each side of the system. Our results highlight the importance of the substrate structure on thermal resonance and could inspire further developments in designing thermal devices.
\end{abstract}
%}

%\noindent{Keywords: transport processes; heat conduction; coupled FK lattices}

% 44.10.+i Heat conduction
% 05.60.-k Transport processes
% 05.45.-a Nonlinear dynamics and chaos
% 05.10.Gg Stochastic analysis methods

\maketitle

%%%%%%%%%%%%%%%%%%%%%%%%%%%%%%%%%%%%%%%%%%%%%%%%%%%%%%%%%%
\section{INTRODUCTION\label{sec:Intro}}
%%%%%%%%%%%%%%%%%%%%%%%%%%%%%%%%%%%%%%%%%%%%%%%%%%%%%%%%%%

With the shrinkage of electronic devices down to the nanoscale~\cite{Yan11,Lu07} and the increasing interest in thermoelectrics~\cite{Yan22}, the thermal transport properties of nanomaterials have attracted much attention in recent years. In particular, with the emergence of single-molecule electronics~\cite{Cuevas10,Xuefeng14,Su16}, where long-chain molecules attached to tiny electrodes are used to transport and switch electrons, the related problem of thermal transport through molecular junctions has received considerable theoretical and experimental attention. When an electron is transported through a molecule, a fraction of its kinetic energy is transformed into molecular vibrational energy. It has been estimated that 10\% to 50\% of the electron's energy could be converted to heat, so that a power of $10^{11}$ eV/s may be dissipated on a molecular electronic bridge carrying 10 nA under a bias of 1 eV~\cite{Segal03}. The ensuing temperature jump can affect the stability and integrity of the molecular junction. Therefore the rate at which heat is transported away from the conducting junction is crucial to the successful realization of such nanoscale electronic devices. One way to attain this goal is to employ a temporal modulation to direct heat from one part of the device to another or to a thermal reservoir by means of an applied external work. Such a strategy has been tested in a simplified model, the so-called Frenkel-Kontorova (FK) model~\cite{Braun04} ---which is a one-dimensional (1D) harmonic lattice affected by an onsite potential. Previously a 1D model consisting of two dissimilar FK lattices connected together by a time-modulated harmonic coupling under the influence of a static thermal bias was considered, showing that the overlap or separation of the phonon bands associated with the dissimilar segments determines the appearance or absence of thermal resonance (TR), that is, the maximization of a heat flux directed toward the thermal reservoirs obtained for a specific frequency of the external periodic driving~\cite{Bao-quan10,Cuansing10,Zhang11,Romero20}. Thus this proposal offers an option to solve the problem of carrying away the heat from the junction between two dissimilar systems.

Since the onsite FK potential takes into account, in a very simplified way, the influence of a substrate on the thermal conductivity of the harmonic lattice, it is important to recall some previous work on the effects of coupling to substrates on thermal conductivity. It is known that, in certain regions, the negative effect of phonon scattering can be suppressed and the thermal conductivity of nanomaterials can be significantly increased due to the coupling-induced shift of the phonon band to the low-wave vector; thus the thermal conductivity can be efficiently manipulated via coupling to different substrates without altering the conducting structures. The systems so far considered are an anharmonic oscillator chain, a modified double-walled carbon nanotube~\cite{Guo11}, and, very recently, Tungsten ditelluride flakes coupled to various substrates~\cite{Laxmi23}. Also in an study of thermal transport in graphene supported on a silicon dioxide substrate it was determined that increasing the strength of the graphene-substrate interaction enhances the heat flow and effective thermal conductivity along supported graphene, contrary to expectations~\cite{Ong11}. Very recently it was determined that the higher phonon group velocity and relaxation time caused by the weak phonon anharmonicity lead to higher thermal conductivity of gallium indium antimonide alloys~\cite{Zhu23}. It is also known that heat transfer in multicomponent nonmetallic and semiconductor nanomaterials is dramatically affected by scattering of phonons at surfaces and interfaces~\cite{Varshney12}. Therefore, by simulating heat transfer in nanostructures based on a model of FK lattices interacting via van der Waals interactions it was determined that such interface interaction can adjust the thermal conductivity parallel to the interface; this result agrees well with experimental results for carbon nanotube bundles, multiwalled carbon nanotubes, multilayer graphene, and nanoribbons~\cite{Sun13}. It is then clear that the detailed structure of the substrate can have a significant influence on heat transport properties of the attached material. Therefore we expect that this is also the case for the TR phenomenon studied by means of the dissimilar FK lattice previously mentioned. 

In this work we begin to address the influence of substrate structure on the TR phenomenon previously studied in the dissimilar FK lattice by considering the effect of the potential period on the aforementioned phenomenon. The onsite potential in the FK model plays the role of the substrate. Therefore, a modification of the period amounts to considering, in a simplified way, the influence of the substrate on TR. The previously considered setup for the study of TR in the FK model~\cite{Romero20} corresponds to the commensurate case where all oscillators occupy the minima of the substrate potential. In this case the lattice constant and the potential period have the same value and thus did not appear explicitly in the model. However, in the more general considered case the FK possesses two competing length scales. The interoscillator interaction favors an equidistant separation between oscillators, whereas the interaction with the substrate tends to force the oscillators into a configuration in which they are regularly spaced at a distance that corresponds to the period of the external potential. It is then expected that the aforementioned competition of length scales has a significant effect on the TR phenomenon previously observed in the dissimilar FK lattice.

This paper is organized as follows: in Sec.~\ref{sec:Model} the model system and methodology are presented. Our results on the dependence of the TR on the period of the FK potential as well as other structural parameters of the model are reported in Sec.~\ref{sec:Res}. The discussion of the results and our conclusions are presented in Sec.~\ref{sec:Disc}.

%%%%%%%%%%%%%%%%%%%%%%%%%%%%%%%%%%%%%%%%%%%%%%%%%%%%%%%%%%%%%%%%%%%%%%%%
\section{THE MODEL\label{sec:Model}}
%%%%%%%%%%%%%%%%%%%%%%%%%%%%%%%%%%%%%%%%%%%%%%%%%%%%%%%%%%%%%%%%%%%%%%%%

\begin{figure}\centering
\includegraphics[width=0.85\linewidth,angle=0.0]{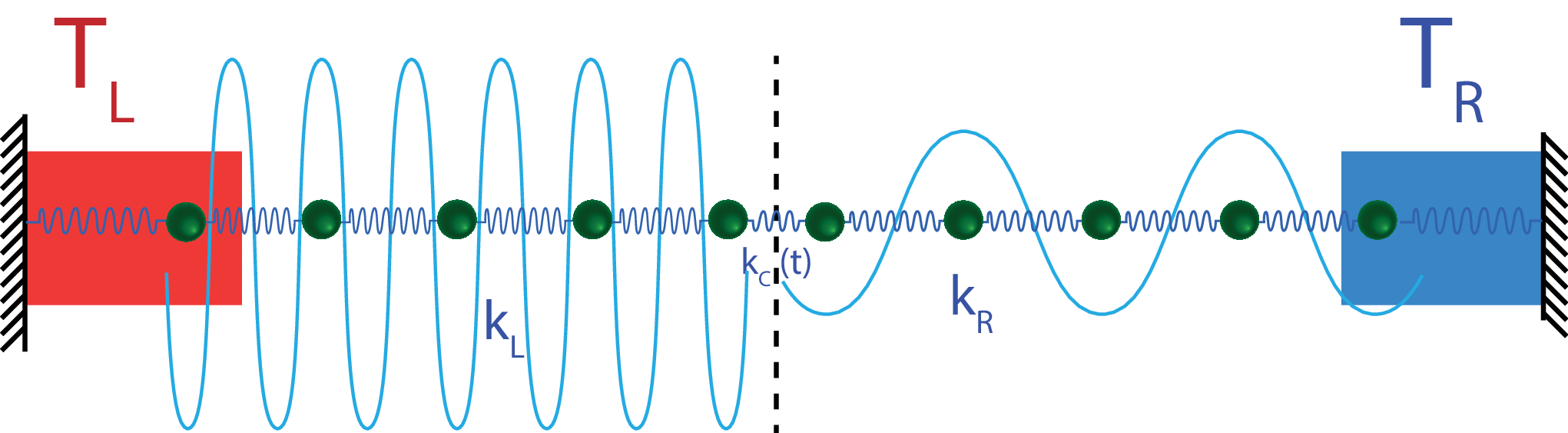}
\caption{Sketch of our model system composed of two dissimilar Frenkel-Kontorova lattices connected by a time-modulated harmonic interaction and in contact with two thermal reservoirs. Periods of Frenkel-Kontorova potential in both sides are different.}
\label{fig:1}
\end{figure}

Our system consists of two dissimilar anharmonic 1D lattices ($L,R$) coupled together by a harmonic spring with a time-modulated strength $\kc(t)$ as shown in Fig.~\ref{fig:1}. The equations of motion (EOM) for a given oscillator within the ensuing size $N$ lattice can be written, in terms of dimensionless variables, as $\dot q_i =p_i/m_i$ and
\ba
\dot p_i&=&k_{i-1}(q_{i-1} - q_i) + k_{i}(q_{i+1} - q_i) - {V_{i}\over2\pi}\sin\bigg(\!2\pi{x_i\over b_{i}}\!\bigg) \cr
   & + & (\xi_{_1} - \gamma_{_1} p_i)\,\delta_{i1} + (\xi_{_N} - \gamma_{_N} p_i)\,\delta_{i{N}},
\ea
where $x_i = q_i + ia$---being $a=1$ the lattice spacing---and
\ba
k_i&=&\sum_{j=1}^{\nc-1}\kl\delta_{ij} + \kc(t)[\delta_{i \nc}+\delta_{i(\nc+1)}] +\!\!\!\! \sum_{j=\nc+2}^N\!\!\!\kr\delta_{ij}, \cr
V_i&=&\sum_{j=1}^{\nc}\Vl\delta_{ij} + \!\!\sum_{j=\nc+1}^N\!\!\Vr\delta_{ij}, \cr
b_i&=&\sum_{j=1}^{\nc}\bl\delta_{ij} + \!\!\sum_{j=\nc+1}^N\!\!\br\delta_{ij}.
\ea
In the above equations $k_{_{L,R}}$, $V_{_{L,R}}$, and $b_{_{L,R}}$ are the harmonic spring constant, amplitude, and period of the FK onsite potential in each segment, respectively; for the rest of the paper we fix the value of $\nc=N/2$. To reduce the number of adjustable parameters we set $\Vr=\lambda\Vl$ and $\kr=\lambda\kl$ and in the following we fix the values of $\Vl=5$ and $\kl=1$. $\{m_i,q_i,p_i\}_{i=1}^{N}$ are the dimensionless mass, displacement, and momentum of the $i$th oscillator; the procedure to construct such dimensionless variables can be seen in the Appendix of Ref.~\cite{Li12}. Fixed boundary conditions are assumed ($q_{_0}=q_{_{{N}+1}}=0$). Henceforth we consider a homogeneous system, i.e. $m_i=1\,\,\forall\,i$. The Gaussian white noise $\xi_{_{1,N}}$ has zero mean and correlation $\langle\xi_{_{1,N}}(t)\xi_{_{1,N}}(t^{\prime})\rangle=2\gamma_{_{1,N}}T_{_{1,N}}m_i(\delta_{1i}+\delta_{{N}i})\delta(t-t^{\prime})$, being $\gamma_{_{1,N}}$ (taken as $0.5$ in all computations hereafter reported) the coupling strength between the system and the left and right thermal reservoirs operating at temperatures $T_{_L}=0.15$ and $T_{_R}=0.05$, respectively; thus the average temperature of the system is $T_{_0}\equiv(T_{_L}+T_{_R})/2=0.1$. Both segments of the system are coupled by means of a time-modulated harmonic coupling of magnitude $\kc(t)=\ko (1+\sin\omega t)$, which is an external driving with frequency $\omega$. The EOM were integrated with a stochastic velocity-Verlet integrator with a timestep of $5\times10^{-3}$ for a stationary time interval of $2\times10^7$ time units after a transient time of $10^{8}$ time units.

%%%%%%%%%%%%%%%%%%%%%%%%%%%%%%%%FIG. 1%%%%%%%%%%%%%%%%%%%%%%%%%%%%%%%%%%%%%%%%%%%%%%
%%%%%%%%%%%%%%%%%%%%%%%%%%%%%%%%FIG. 1%%%%%%%%%%%%%%%%%%%%%%%%%%%%%%%%%%%%%%%%%%%%%%

Once the non-equilibrium stationary state is attained, the local heat flux is computed as ${J}_i=k_i\langle\dot q_{i}(q_{i}-q_{i+1})\rangle$i, with $k_i=\kl$ if $i\in[2,\nc-1]$, $k_i=\kc(t)$ if $i=\nc$, and $k_i=\kr$ if $i\in[\nc+1,N]$, and the local temperature at each site as $T_i=\langle p_i^2/m_i\rangle$; in both instances $\langle\cdots\rangle$ indicates time average over the entire time interval corresponding to the stationary state. In the latter the heat flux within each segment becomes independent of the site. Thus, to improve the statistical precision of our results, the mean heat flux $J_{_{L,R}}$ on each side of the lattice is calculated as the algebraic average of ${J}_i$ over the number of unthermostatted oscillators in each segment. Now, the rate of work $\dot W$ done by the external driving in the contact at $\nc$ is dissipated into the reservoirs, implying that $\dot W = \Jl + \Jr$, where $J_{_{L,R}}$ are defined as positive when the heat flows into the reservoirs.

%%%%%%%%%%%%%%%%%%%%%%%%%%%%%%%%%%%%%%%%%%%%%%%%%%%%%%%%%%%%%%%%%%%%%%%%%%%%%%%%%%%%%%
\section{RESULTS\label{sec:Res}}
%%%%%%%%%%%%%%%%%%%%%%%%%%%%%%%%%%%%%%%%%%%%%%%%%%%%%%%%%%%%%%%%%%%%%%%%%%%%%%%%%%%%%%

\subsection{Equal FK period}

\begin{figure}\centering
\includegraphics[width=0.95\linewidth,angle=0.0]{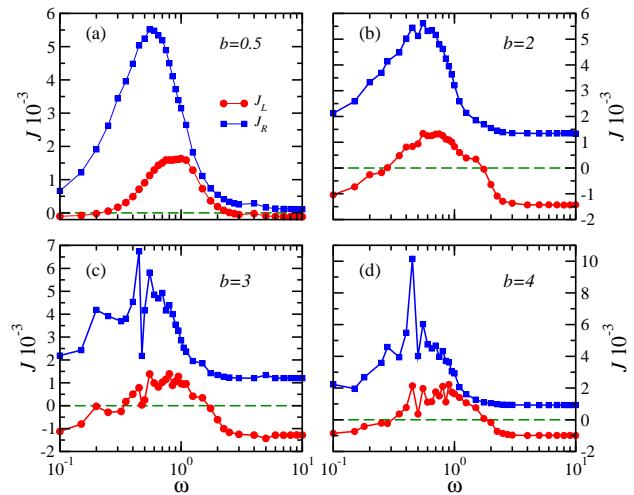}
\caption{Heat flux versus driving frequency $\omega$ for various values of the period of the Frenkel-Kontorova potential $b\equiv b_{_L}=b_{_R}$. The energy currents through the left and right segments are $\Jl$ and $\Jr$, respectively. Parameters are $\lambda=0.2$, $\ko=0.05$, and $N=32$. Continuous lines are a guide to the eye.}
\label{fig:2}
\end{figure}

In Fig.~\ref{fig:2} we present the results of the dependence of heat fluxes $\Jl$ and $\Jr$ as a function of the driving frequency $\omega$ with $\lambda=0.2$, $\ko=0.05$ and $N=32$ in the case without a mismatch of the FK periods on both sides of the system by setting $b_{_L}=b_{_R}\equiv b$. In the adiabatic driving limit $\omega\rightarrow0$ the heat flows from left to right of the system, with $\Jr=-\Jl>0$, and thus the averaged net power released to the system is zero. In the opposite limit $\omega\rightarrow\infty$ the coupling oscillates very fast and converges to a time average constant value $\ko$ as if there is no driving. Thus it can be readily observed that, for all depicted cases in Fig.~\ref{fig:2}, the relevant phenomenology occurs within the intermediate range $\omega\in[0.1,2]$ of the driving frequency, where the power afforded by the external driving and released in the contact region is dissipated into the reservoirs, since now $\Jl>0$. This phenomenon, due to the resonant interaction of the external drive with the system's intrinsic frequencies, has been known for a time as TR as already mentioned in the Introduction. For the previously considered case wherein $b_{_L}=b_{_R}=1$~\cite{Romero20} the average $(\Jl+\Jr)/2$ was considered, with its value maximized at a specific driving frequency value. However, in Fig.~\ref{fig:2} it can be noted that, except for $b=2$, the maxima of each heat flux are obtained at significantly different driving frequencies. It is clear that, as $b$ grows, the maximum of $\Jr$ also does so and becomes increasingly defined whereas the one for $\Jl$ diminishes in magnitude, with the resonant frequency of the former being lower than that of the latter, except for $b=2$ as mentioned previoulsy.

%%%%%%%%%%%%%%%%%%%%%%%%%%%%%%%%IG. 2%%%%%%%%%%%%%%%%%%%%%%%%%%%%%%%%%%%%%%%%%%%%%%
%%%%%%%%%%%%%%%%%%%%%%%%%%%%%%%%FIG. 2%%%%%%%%%%%%%%%%%%%%%%%%%%%%%%%%%%%%%%%%%%%%%%

\begin{figure}\centering
\includegraphics[width=0.85\linewidth,angle=0.0]{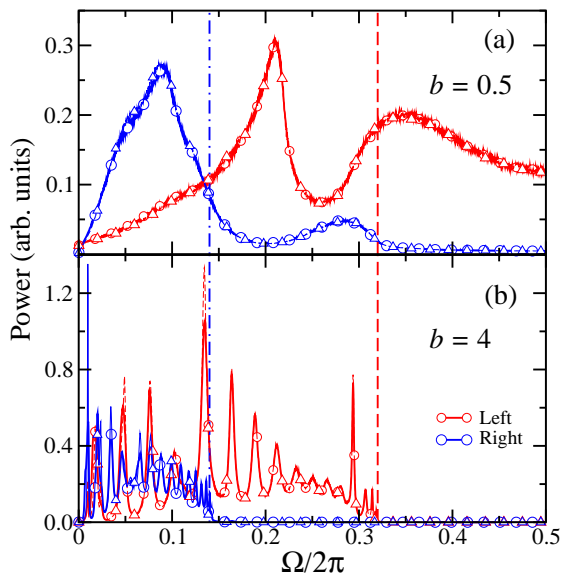}
\caption{(a) Power spectra of oscillators $i=16$ and $i=17$ in the left (red) and right (blue) sides of the contact for the case of $b=0.5$ depicted in the previous figure. Circles correspond to a driving frequency of $\omega=10^{-3}$ and triangles to $\omega=10$. (b) Same as (a) but for $b=4$. Vertical dashed and dot-dashed lines correspond to the cut-off spectral frequencies of the left and right phonon bands, respectively. Same $\lambda$, $\ko$, and $N$ values as in previous figure.}
\label{fig:3}
\end{figure}

To understand the origin of these mentioned features of the system we first plot in Fig.~\ref{fig:3} the phonon spectra (PS) $P(\Omega)=\langle|\tau^{-1}\!\!\int_{_0}^{\tau}\!\! dt\dot q_i(t)\exp(-\mathrm{i}\Omega t)|^2\rangle$---computed over an interval of $\tau=10^{13}$ time units---of the interface oscillators at the left and right side of the contact ($i=16$ and $17$ for the considered $N$ value) for the two extreme values of the FK period $b$ reported in Fig.~\ref{fig:2} and, for each case, two values of the external drive, $\omega=10^{-3}$ and $10$, that correspond to the adiabatic and high-frequency driving limits, respectively. These limits correspond to an energy transport regime without external driving, which explains why the spectra for the two $\omega$ values overlap almost perfectly. Furthermore, for both depicted instances of the period $b$ there is a spectra overlap in the low $\Omega$ frequency range. Thus the energy transport goes through the phonon channels determined by the thermal bias imposed at the boundaries.

The sizable power that the spectra in Fig.~\ref{fig:3} have within the low spectral frequency range $\Omega$ can be understood if we recall that there is a critical temperature $T_{_{\mathrm{cr}}}\approx V/(2\pi)^2$ above which the kinetic energy is large enough to overcome the onsite potential barrier, where the contribution of the latter can be neglected~\cite{Li04a}. We consider the temperatures of the boundary oscillators $T_{_{16}}$ and $T_{_{17}}$ to assess the dynamical regime in which each side is. In our particular case we have $T_{_{\mathrm{cr}}}^{(L)}=0.1266\lesssim T_{_{16}}\approx0.1267$ for $V_{_L}=5$ and $T_{_{\mathrm{cr}}}^{(R)}=0.025<T_{_{17}}\approx0.05$ for $V_{_R}=1$. Although the temperature in the left boundary is only slightly above the corresponding critical temperature, it can be readily seen in the figure that the spectrum corresponding to the left side is populated in the low-frequency range, a characteristic that indicates that the influence of the left onsite potential is already very weak. Thus the presented evidence indicates that both sides of the system are in a temperature regime wherein they behave as harmonic lattices with a phonon band of $0<\Omega<({4k_{_{L,R}}})^{1/2}$ composed mainly of noninteracting low-frequency phonons, which gives $0<\Omega/2\pi\lesssim0.32$ for the left oscillator and $0<\Omega/2\pi\lesssim0.14$ for the right. Now, thermal transport can, in principle, occur only for frequency values in the overlapping region of these phonon bands, which implies the associated phonons with the process must have frequencies within the range $\Omega/2\pi\lesssim0.14$. It is also to be noted that the FK period value $b$ has a strong influence of the morphology of the phonon bands. For the case $b=0.5$ depicted in Fig.~\ref{fig:3}(a) the structure of the substrate, and therefore the amplitude of the FK potential, is relevant since there is a sizable amount of high-frequency phonons in the spectrum associated with the side of the system in contact with the hot reservoir, which contrast with that in contact with the cold one, that has most of its spectral power concentrated in the low-frequency range. By contrast, the spectra depicted in Fig.~\ref{fig:3}(b) with $b=4$ present a more discontinuous structure that signals an underlying harmonic dynamics. This feature is explained by the fact that, for such large value of the FK period, the length scale associated with the substrate is way larger than that of the coupled FK lattice, and thus the influence of the underlying substrate is negligible on the latter. This leads to the observed harmonic-like structure of the spectra. Now, the power of the spectra in panel (b) is larger than that of the ones depicted in panel (a). This fact, together with the sizable overlap in the low-frequency region, accounts for the increase in value of $\Jr$ form $b=0.5$ to 4 observed in Fig.~\ref{fig:2}. On the contrary, since the left spectrum in (b) has sizable power allocated in intermediate frequencies, then $\Jl<\Jr$ for $b=4$.

%%%%%%%%%%%%%%%%%%%%%%%%%%%%%%%%FIG. 3%%%%%%%%%%%%%%%%%%%%%%%%%%%%%%%%%%%%%%%%%%%%%%
%%%%%%%%%%%%%%%%%%%%%%%%%%%%%%%%FIG. 3%%%%%%%%%%%%%%%%%%%%%%%%%%%%%%%%%%%%%%%%%%%%%%

\begin{figure}\centering
\includegraphics[width=0.85\linewidth,angle=0.0]{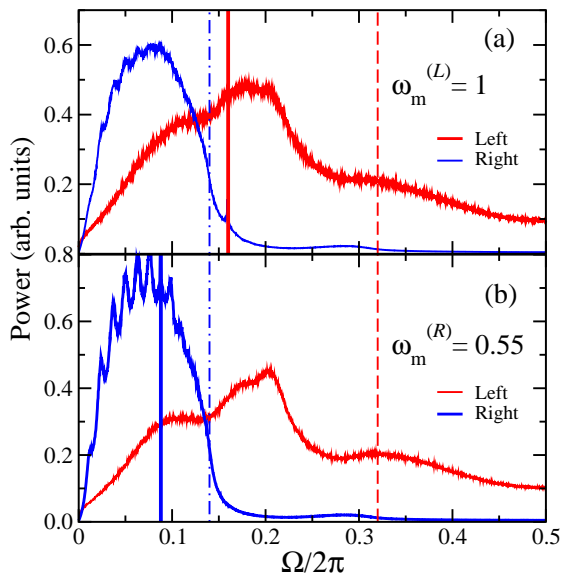}
\caption{(a) Power spectra of oscillators $i=16$ and $i=17$ in the left (red) and right (blue) sides of the contact for a lattice with period $b=0.5$ and a resonant frequency of $\Oml=1$. (b) Same as (a) but for $\Omr=0.55$. Vertical dashed and dot-dashed lines correspond to the cutoff spectral frequencies of the left and right phonon bands, respectively. Vertical solid lines denote the value $\Oml/2\pi$ in (a) and $\Omr/2\pi$ in (b). Same $\lambda$, $\ko$, and $N$ values as in previous figure.}
\label{fig:4}
\end{figure}

Next, in Fig.~\ref{fig:4} we again plot the PS of the oscillators to the left and right of the contact for a period of $b=0.5$, but now in the thermal resonance regime; for the left side the resonant frequency is $\Oml=1$ and for the right, $\Omr=0.55$, as can be observed from panel (a) of Fig.~\ref{fig:2}. We first notice that the spectra in both panels of Fig.~\ref{fig:4} have almost twice the spectral power than those corresponding to the non-resonant frequency regimes depicted in Fig.~\ref{fig:3}(a), with a significant amount of spectral power of the left spectra shifted toward lower spectral frequencies compared with the undriven case. For the right spectra the contribution of intermediate-frequency phonons has been completely suppressed, with the spectral power almost entirely concentrated within the low-frequency range. Now, examining the low-frequency region of Fig.~\ref{fig:4}(a) we notice that, in the overlapping frequency range of both phonon bands, the left side spectrum increases monotonically, which skews the maximum spectral power obtained from the contribution of both spectra toward higher-frequency values, thus rendering a resonant driving frequency for the left side of the system of $\Oml/2\pi=0.16$, which lies outside the overlapping region $\omega_{\mathrm{m}}^{_{(L,R)}}/2\pi\lesssim0.14$. It is worth noting that $\Oml$ corresponds to the lowest peak in Fig.~\ref{fig:2}(a), which has the lowest contribution to the overall TR effect observed in the aforementioned figure. In Fig.~\ref{fig:4}(b) we notice there is more spectral power in the right side spectrum than in the left-side one, and even the overall contribution of the latter is slightly lower than that of the corresponding one in the previous panel. These factors explain the higher values of $\Jr$ compared with $\Jl$ ones in the intermediate driving frequency range that can be observed in Fig.~\ref{fig:2}. Now, the obtained value of $\Omr/2\pi=0.09$---which lies within the overlapping region and corresponds to the maximum peak observed in Fig~\ref{fig:2}(b)---can be readily explained by noting that the left-side spectrum, contrary to the corresponding one in the previous panel, is nonmonotonic in the overlapping frequency region; thus $\Omr$ should be shifted towards lower frequency values and closer to the value corresponding to the highest peak of the right-side spectrum. At this point it is important to note that, in both panels of Fig.~\ref{fig:4}, there is a small spike-like perturbation on the right-side spectra that almost coincides with the $\omega_{\mathrm{m}}^{_{(L,R)}}/2\pi$ values displayed. This fact could be evidence that the spectral frequencies associated with the right side are the ones that ultimately control the resonant frequency values, which in hindsight could be expected since that is the one with the lowest amplitude of the FK potential, thus allowing the low-frequency band to be more populated, which most heavily controls the heat flux through the system.

%%%%%%%%%%%%%%%%%%%%%%%%%%%%%%%%FIG. 4%%%%%%%%%%%%%%%%%%%%%%%%%%%%%%%%%%%%%%%%%%%%%%
%%%%%%%%%%%%%%%%%%%%%%%%%%%%%%%%FIG. 4%%%%%%%%%%%%%%%%%%%%%%%%%%%%%%%%%%%%%%%%%%%%%%

\begin{figure}\centering
\includegraphics[width=0.85\linewidth,angle=0.0]{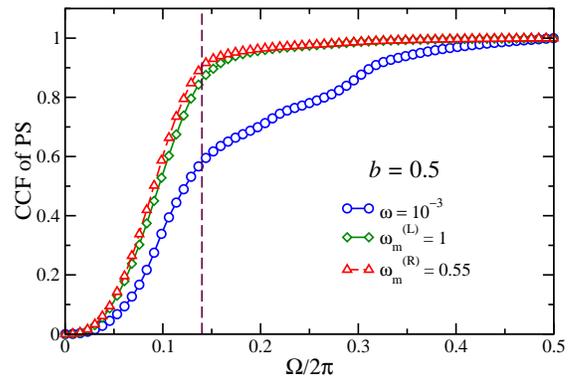}
\caption{Cumulative correlation factor of phonon spectra between oscillators $i=16$ and $17$ for $\Om$ values corresponding to non-driven and driven regimes of Figs.~\ref{fig:3}(a) and \ref{fig:4} respectively. Vertical dashed line indicates the spectral frequency region $\Omega/2\pi\lesssim0.14$ where the phonon spectra in Fig.~\ref{fig:4} overlap. Same $\lambda$, $\ko$, $b$, and $N$ values as in Fig.~\ref{fig:2}.}
\label{fig:5}
\end{figure}

To quantify in more detail the degree of overlap of the power spectra between the boundary oscillators, and thus gain further insight into the mechanisms responsible for TR, the cumulative correlation factor (CCF), introduced in Refs.~\cite{Li05,Zhang17}, is used to represent the match-mismatch degree of vibrational modes among them. The CCF below a specific frequency $\omega_s$ between the PS of oscillators in the left and right sides of the interface is defined as
\be
M_{_{L,R}}(\omega_s)={\int_0^{\omega_s}P_{_{L}}(\omega)P_{_{R}}(\omega)d\omega\over\int_0^{\infty}P_{_{L}}(\omega)d\omega\int_0^{\infty}P_{_{R}}(\omega)d\omega}.
\ee
Each CCF is normalized by dividing $M(\omega_s)$ by $M(\infty)$. Previously it has been established that a large CCF value at the low frequency range means lower mismatch, resulting in a higher heat flux~\cite{Dong19}. Therefore it is reasonable to infer that the CCF can shed some light onto the mechanism at the origin of TR. In Fig.~\ref{fig:5} we present the CCF for the $\Om=10^{-3}$ case depicted in Fig.~\ref{fig:3}(a), and for the $\Oml$ and $\Omr$ instances of Fig.~\ref{fig:4}. It is clear that, within the resonant frequency regime $\Omega/2\pi\lesssim0.14$ previously determined, the CCFs corresponding to the resonant frequency values $\Oml$ and $\Omr$ are higher than the CCF corresponding to the nondriven regime of $\Om=10^{-3}$. This indicates a higher phonon match and thus a higher heat flux value in the former regime than in the latter one. Furthermore, since the curves for $\Oml$ and $\Omr$ are virtually identical, it indicates that the same mechanism, that is, TR, is responsible for their morphology. Thus all these factors are consistent with the picture that the overlap of the phonon bands plays a crucial role in the appearance of the TR phenomenon, since the resonant frequency has to have a value within or very close to the region where the aforementioned phonon bands overlap.

%%%%%%%%%%%%%%%%%%%%%%%%%%%%%%%%FIG. 5%%%%%%%%%%%%%%%%%%%%%%%%%%%%%%%%%%%%%%%%%%%%%%
%%%%%%%%%%%%%%%%%%%%%%%%%%%%%%%%%%%%%%%%%%%%%%%%%%%%%%%%%%%%%%%%%%%%%%%%%%%%%%%%%%%%

\begin{figure}\centering
\includegraphics[width=0.75\linewidth,angle=0.0]{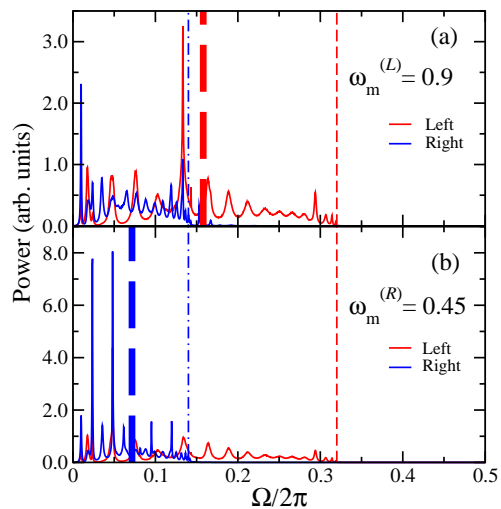}
\caption{(a) Power spectra of the two oscillators in the left (red) and right (blue) sides of the contact for a lattice with period $b=4$ and a resonant frequency of $\Oml=0.9$. (b) Same as (a) but for $\Omr=0.45$. Vertical dashed and dot-dashed lines correspond to the cutoff spectral frequencies of the left and right phonon bands, respectively. Vertical thick dashed lines denote the value $\Oml/2\pi$ in (a) and $\Omr/2\pi$ in (b). Same $\lambda$, $\ko$, and $N$ values as in previous figure.}
\label{fig:6}
\end{figure}

The PS for the boundary oscillators corresponding to the resonant regime and a FK period of $b=4$ are presented in Fig.~\ref{fig:6}. For the lowest and least defined peak of $\Jl$ versus $\omega$ in Fig.~\ref{fig:2}(d), the left spectrum depicted in Fig.~\ref{fig:6}(a) presents a large single peak that skews the resonant frequency of $\Oml/2\pi=0.143$ just outside the overlapping range $\Omega/2\pi\lesssim0.14$, as in the $b=0.5$ already presented in Fig.~\ref{fig:4}(a). However, for the very definite peak of $\Jr$ vs $\omega$ observed in Fig.~\ref{fig:2}(d) the corresponding spectra presented in Fig.~\ref{fig:6}(b) reveal that two single, very low-frequency peaks in the right spectrum drive the resonant frequency $\Omr=0.45$ within the overlapping region. These results seem to suggest that, strictly speaking, the explanation of the TR effect in terms of the overlapping phonon bands can only be associated with certainty to a large, well-defined peak in the dependence of the heat flux on the external driving frequency. Therefore we guess that a modification to accommodate the effect of the substrate on the phonon band description is needed.

%%%%%%%%%%%%%%%%%%%%%%%%%%%%%%%%FIG. 6%%%%%%%%%%%%%%%%%%%%%%%%%%%%%%%%%%%%%%%%%%%%%%
%%%%%%%%%%%%%%%%%%%%%%%%%%%%%%%%FIG. 6%%%%%%%%%%%%%%%%%%%%%%%%%%%%%%%%%%%%%%%%%%%%%%

\begin{figure}\centering
\includegraphics[width=0.85\linewidth,angle=0.0]{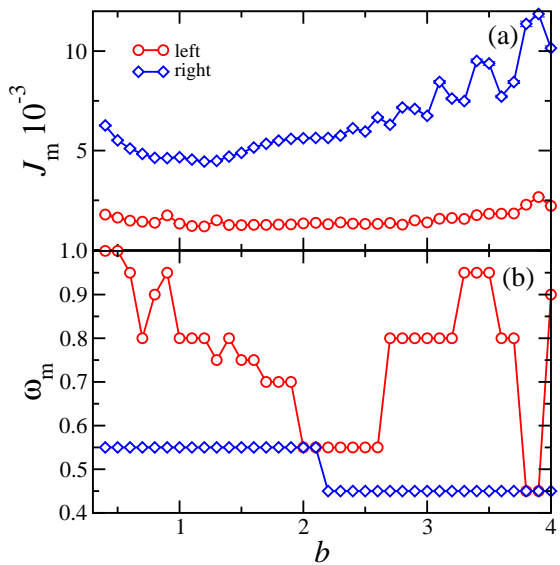}
\caption{(a) Maximum heat flux $\Jm$ in the left (circles) and right (diamonds) sides of the system versus Frenkel-Kontorova period $b$. (b) Resonant frequencies $\Om$ corresponding to the $\Jm$ values of the previous panel versus Frenkel-Kontorova period $b$. Same parameter values as in Fig.~\ref{fig:2}. Continuous lines are a guide to the eye.}
\label{fig:7}
\end{figure}

To complement the information presented in Fig.~\ref{fig:2} we report the behavior of the maximum value of the heat fluxes in each side of the system as a function of the FK period $b$ in Fig.~\ref{fig:7}(a). It is evident the almost monotonic increase of the $\Jm$ value for the right side and the weak dependence of the corresponding value on the left side as $b$ increases. The behavior of the resonant frequencies for each side as a function of $b$, presented in panel (b), indicates that a sharp but small decrease in $\Om$ of the right side around $b\sim2$ correlates with a significant, albeit nonmonotonic, increase of the corresponding $\Jm$ value. The only observable correlation of the behavior of $\Om$ in the left side with the associated $\Jm$ value is that the sudden variation of the resonant frequency that appears at $b=2.5$ with a very slight increase of $\Jm$ in the left side observed in panel (a) for the same $b$ value range.

%%%%%%%%%%%%%%%%%%%%%%%%%%%%%%%%FIG. 7%%%%%%%%%%%%%%%%%%%%%%%%%%%%%%%%%%%%%%%%%%%%%%
%%%%%%%%%%%%%%%%%%%%%%%%%%%%%%%%FIG. 7%%%%%%%%%%%%%%%%%%%%%%%%%%%%%%%%%%%%%%%%%%%%%%

\begin{figure}
\centerline{\includegraphics*[width=75mm]{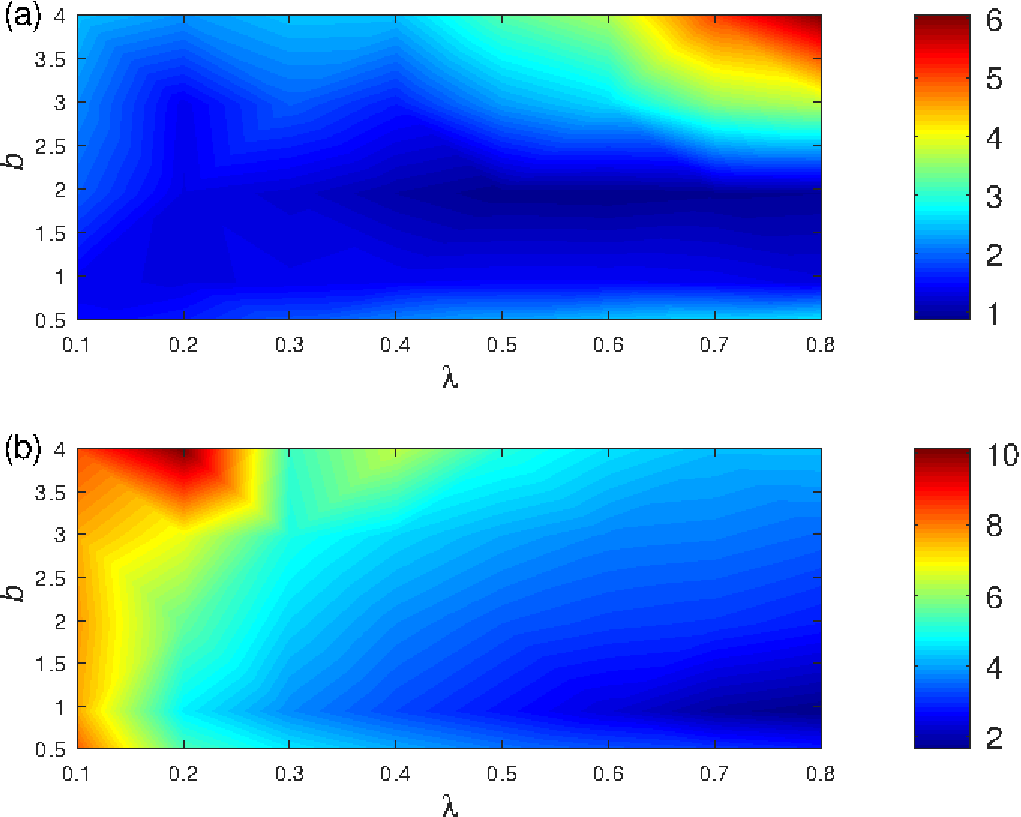}}
\caption{Contour plot of the maximum heat flux $\Jm(\lambda,b)\times10^{-3}$ in the (a) left and (b) right segments of the system for $\ko=0.05$ and $N=32$.}
\label{fig:8}
\end{figure}

\begin{figure}\centering
\includegraphics[width=0.85\linewidth,angle=0.0]{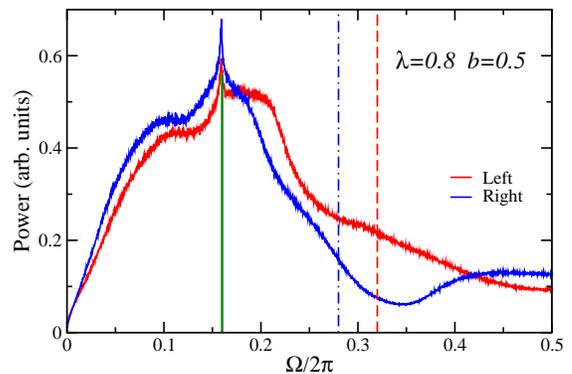}
\caption{Power spectra of the two oscillators in the left (red) and right (blue) sides of the contact for a lattice with Frenkel-Kontorova period of $b=0.5$, $\lambda=0.8$, and a single resonant frequency of $\Oml=\Omr=1$. Vertical dashed and dot-dashed lines correspond to the cutoff spectral frequencies of the left and right phonon bands, respectively. Vertical solid line denotes the $\Om/2\pi$ value. Same $\ko$, and $N$ values as in previous figure.}
\label{fig:9}
\end{figure}

Since the structural asymmetry is controlled with the parameter $\lambda$, it is important to study the dependence of the TR phenomenon upon its variation thereof. Thus, in Fig.~\ref{fig:8} we present the contour plot of the maximum heat flux $\Jm(\lambda,b)\times10^{-3}$ on each side of the system as a function of both the FK period and $\lambda$. It is clear that the phenomenology reported in Fig.~\ref{fig:7}(a) is drastically modified as the asymmetry decreases, that is, as $\lambda$ increases in value. For the left side it is observed that the maximum $\Jm$ are obtained for both large FK period $b$ and $\lambda$ values, whereas for the right one those are obtained with small values of $\lambda$ and large $b$ values. It is also clear that the value range for $\Jm$ in the right side is almost double in size compared with that in the left one. This result can be accounted for by the fact that, for small $\lambda$ values, $\Vr$ is also small, and in this case energy transports superdiffusively~\cite{Li05a}. Now, for $\lambda=0.8$ and $b=0.5$ the maximum heat fluxes on each sides are of similar magnitude. For the aforementioned instance the corresponding spectra are reported in Fig.~\ref{fig:9}. Since now the system is highly symmetrical, both spectra have very similar morphology, specially in the low-frequency range. The phonon bands are $0<\Omega/2\pi\lesssim0.32$ and $0<\Omega/2\pi\lesssim0.28$ for the left and right sides, respectively. Thus the overlapping region of both bands is greatly enlarged compared to with the case depicted in Fig.~\ref{fig:4}, since now we have $\Om/2\pi\lesssim0.28$. Therefore, most of the spectral power is concentrated in this frequency range, with a resonant frequency of $\Om/2\pi\sim0.1592$ clearly within it. The latter is clearly visible as a spike-like perturbation in both spectra at that $\Om$ value.

%%%%%%%%%%%%%%%%%%%%%%%%%%%%%%%%FIG. 8%%%%%%%%%%%%%%%%%%%%%%%%%%%%%%%%%%%%%%%%
%%%%%%%%%%%%%%%%%%%%%%%%%%%%%%%%FIG. 8%%%%%%%%%%%%%%%%%%%%%%%%%%%%%%%%%%%%%%%%

%%%%%%%%%%%%%%%%%%%%%%%%%%%%%%%%FIG. 9%%%%%%%%%%%%%%%%%%%%%%%%%%%%%%%%%%%%%%%%%%%%%%
%%%%%%%%%%%%%%%%%%%%%%%%%%%%%%%%FIG. 9%%%%%%%%%%%%%%%%%%%%%%%%%%%%%%%%%%%%%%%%%%%%%%

\begin{figure}\centering
\includegraphics[width=0.85\linewidth,angle=0.0]{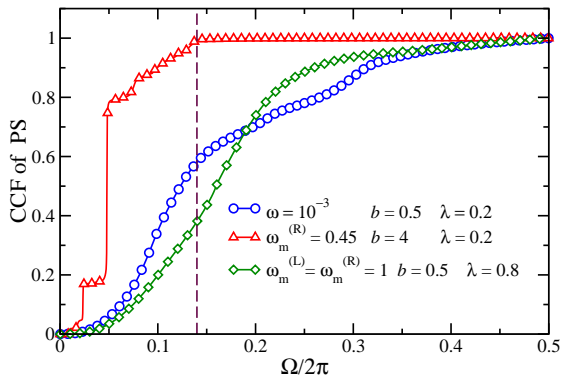}
\caption{Cumulative correlation factor of phonon spectra between oscillators $i=16$ and $17$ for $\Om$ values corresponding to nondriven and driven regimes, the latter with different $b$ and $\lambda$ values. Vertical dashed line indicates the spectral frequency region $\Omega/2\pi\lesssim0.14$ wherein the phonon spectra in Fig.~\ref{fig:4} overlap. Same $\ko$, and $N$ values as in Fig.~\ref{fig:2}.}
\label{fig:10}
\end{figure}

In Fig.~\ref{fig:10} we plot the CCF obtained from the power spectra in Figs.~\ref{fig:6} and~\ref{fig:9}, together with the $\Om=10^{-3}$ instance (undriven case) of Fig.~\ref{fig:3} for reference. It is clear that, for $b=4$, the increase of the CCF in the low-frequency range is much steeper and less monotonic than in the driven cases depicted in Fig.~\ref{fig:5}. Furthermore, for very low frequencies the CCF has a higher value than that of the undriven case. Since this high value means a smaller mismatch, the resulting heat flux and TR effect are stronger than those corresponding to $b=0.5$, as was already noted in Fig.~\ref{fig:7}. Now, for the highly symmetrical case with $b=0.5$ and $\lambda=0.8$ the CCF has a lower value than the one of the undriven case in most of the low-frequency range. This results in a low TR effect, which is, in magnitude, similar to that of the asymmetrical case $\lambda=0.2$ but with $b=4$. Therefore the CCF is a tool that can adequately assess the magnitude of the TR effect from its behavior in the low-frequency regime.

%%%%%%%%%%%%%%%%%%%%%%%%%%%%%%%%FIG. 10%%%%%%%%%%%%%%%%%%%%%%%%%%%%%%%%%%%%%%%%%%%%%
%%%%%%%%%%%%%%%%%%%%%%%%%%%%%%%%%%%%%%%%%%%%%%%%%%%%%%%%%%%%%%%%%%%%%%%%%%%%%%%%%%%%

\subsection{Unequal FK periods}

\begin{figure}
\centerline{\includegraphics*[width=75mm]{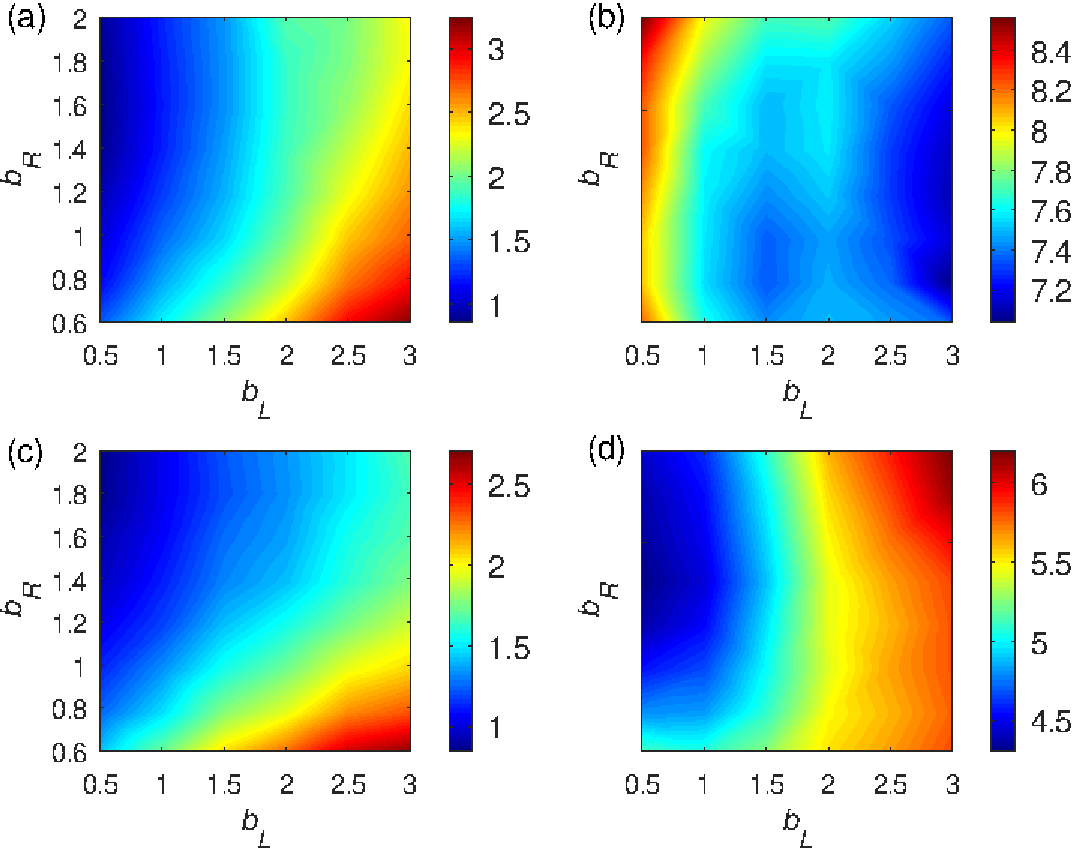}}
\caption{Contour plot of the maximum heat flux $\Jm(\bl,\br)\times10^{-3}$ in the (a) left and (b) right segments of the system for $\lambda=0.1$ in both instances. (c) and (d) are the same as (a) and (b), but for $\lambda=0.2$. Parameters are $\ko=0.05$ and $N=32$.}
\label{fig:11}
\end{figure}

In this section we separately change the period in the two sides of the system, which entails now having $\bl$ and $\br$ as unequal parameters. So, in Fig.~\ref{fig:11} we present the contour plot $\Jm(\bl,\br)\times10^{-3}$ of the maximum heat flux in each segment of the system for the two lowest values of the asymmetry parameter $\lambda$. It can be readily observed that, compared with the case of equal FK periods depicted in Fig.~\ref{fig:8}, the regions in the parameter space wherein $\Jm$ is maximized are greatly modified. More specifically, for $\lambda=0.1$, panels (a) and (b) corresponding to the left and right sides respectively, it can be inferred that, to maximize the heat flux through the left side, the corresponding period has to be in the range $\bl\gtrsim2$, with $\br\lesssim1$. This result has its origin in the fact that, for high $\bl$ values, the left side of the system is largely harmonic, with a ballistic energy transport. On the contrary, $\Jm$ through the right side is maximized in the whole range of $\br$ values, provided that $\bl\approx0.5$. In this case the explanation is again that the low $\Vr$ value ensures a superdiffusive energy transport, irrespective of the FK period value considered~\cite{Li05a}. It is also worth noting the high magnitude disparity of the heat fluxes on both sides. As an example, the combination $\bl=0.5$ and $\br=2$ results in a very large value $\Jm\gtrsim8$ in the right side and a low value $\Jm\sim1$ on the left. For the $\lambda=0.2$ case, panels (c) and (d), the aforementioned disparity of scales is reduced. In the right segment the highest values of $\Jm$ are located in the region bounded by $\bl\gtrsim2$ and $\br\gtrsim1.5$. For this case the increase in the magnitude of $\Vr$ in comparison to the $\lambda=0.1$ instance entails a decrease of the superdiffusive behavior. Therefore now the high $\Jm$ values correspond to high $\bl$ and $\br$, which are associated with ballistic heat transport.

%%%%%%%%%%%%%%%%%%%%%%%%%%%%%%%%FIG. 11%%%%%%%%%%%%%%%%%%%%%%%%%%%%%%%%%%%%%%%
%%%%%%%%%%%%%%%%%%%%%%%%%%%%%%%%FIG. 11%%%%%%%%%%%%%%%%%%%%%%%%%%%%%%%%%%%%%%%

\begin{figure}
\centerline{\includegraphics*[width=75mm]{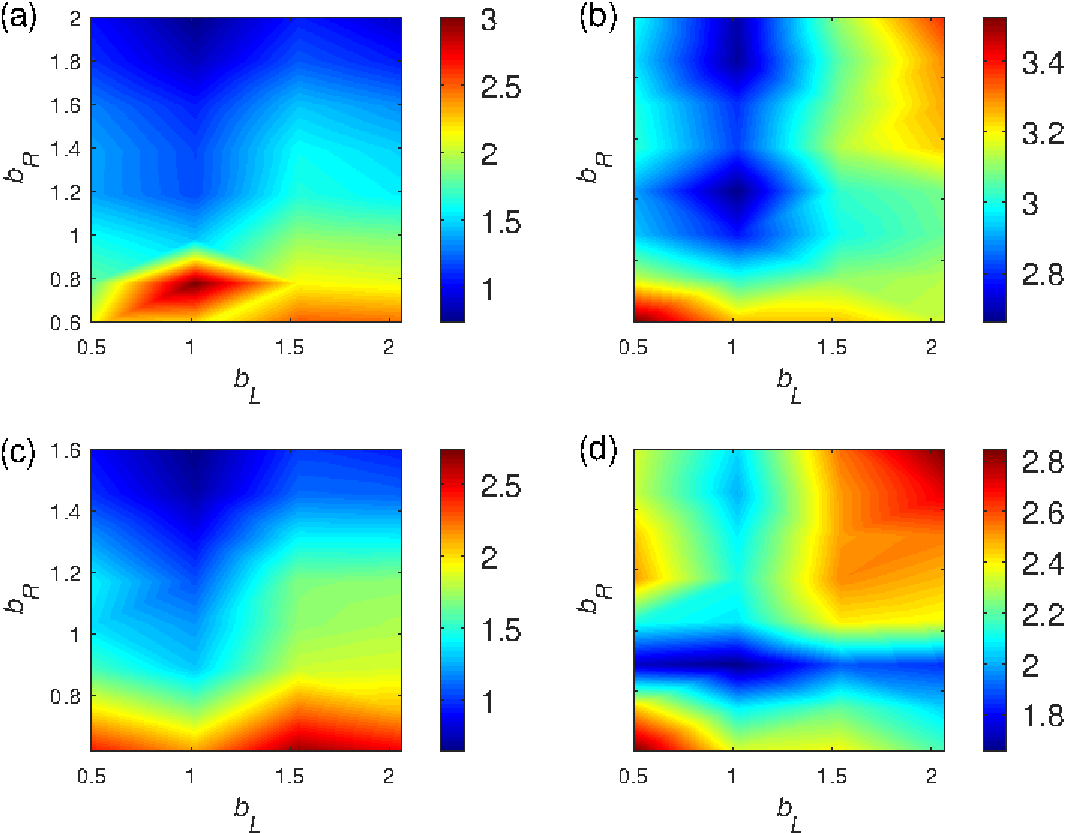}}
\caption{Contour plot of the maximum heat flux $\Jm(b_{_L},b_{_R})\times10^{-3}$ in the (a) left and (b) right segments of the system for $\lambda=0.5$ in both instances; (c) and (d) are the same as (a) and (b), but for $\lambda=0.8$. Parameters are $\ko=0.05$ and $N=32$.}
\label{fig:12}
\end{figure}

The more structurally symmetrical instances of the FK lattice are presented in Fig.~\ref{fig:12}. For $\lambda=0.5$, panels (a) and (b), it is clear that the magnitudes of the heat fluxes on both sides are closer in value than those considered in the corresponding instances of Fig.~\ref{fig:11}. However, the range of FK period values that render the largest $\Jm$ values in both sides is very narrow, being mostly confined to very low $\bl$ and $\br$ values. These are associated with an increase in the influence of the onsite potential, which in turn entails a diffusive energy transport. Finally, for $\lambda=0.8$ the results reported in panels (c) and (d) for the left and right sides indicate that the homogenization in behavior of both sides continues, but in different ways depending of the side of the system. In the left side the maximum $\Jm$ is obtained for $\br\lesssim0.8$ for any $\bl$ value, which has its origin in the fact that, for low $\br$ values, the influence of the $\Vr$ is stronger, thus increasing the diffusive behavior in the left side. On the right side, the large $\Jm$ values for low $\bl$ and $\br$ values have its origin in the diffusive behavior, and for larger ones again in a ballistic behavior.

%%%%%%%%%%%%%%%%%%%%%%%%%%%%%%%%FIG. 12%%%%%%%%%%%%%%%%%%%%%%%%%%%%%%%%%%%%%%%
%%%%%%%%%%%%%%%%%%%%%%%%%%%%%%%%FIG. 12%%%%%%%%%%%%%%%%%%%%%%%%%%%%%%%%%%%%%%%

%%%%%%%%%%%%%%%%%%%%%%%%%%%%%%%%%%%%%%%%%%%%%%%%%%%%%%%%%%
\section{CONCLUDING REMARKS\label{sec:Disc}}
%%%%%%%%%%%%%%%%%%%%%%%%%%%%%%%%%%%%%%%%%%%%%%%%%%%%%%%%%%

The 1D dissimilar FK lattice model wherein the TR phenomenon was previously studied is modified to consider the influence of the period of the FK potential. Such modification allows one to consider, in a simplified way, the influence of the structure of the substrate. Our analysis determined that the maximum magnitude of the resulting heat fluxes is obtained for a period lower than the lattice constant. The maximum values of the heat flux in both sides are obtained for rather different values of the driving frequency, at variance with the commensurate case. Seemingly some of our results can only be partially explained by the overlap of the phonon bands associated with each side of the system, since the resonant frequency of the left side lies outside---albeit very close---to the overlapping region for some of the studied cases. However, it is also important to remark that, since the periodic driving is external to the intrinsic system's dynamics, in principle it is not strictly necessary that the thermal phonons follow exactly the same frequency channels than in the undriven instance. Now, since in the commensurate case the latter was indeed the case, we believe that, although the interaction of the phonon bands certainly lies at the origin of TR, maybe it has to be complemented by some information provided by the anharmonic interaction with the substrate to obtain a more complete picture of this phenomenon when the FK period is relevant. We intend to pursue this study in the near future.

Our study complements those previously performed on the effect of the FK period on heat conduction~\cite{Zhang18} and on its thermal rectification, that is, asymmetric heat flow, properties~\cite{Luo07}. It is interesting to note that the vanishing of the heat flux when the driving frequency exceeds the bandwidth of the lattice with an unitary harmonic constant observed in Fig.~\ref{fig:2} is strikingly similar to the vanishing of the heating rate observed in the transport of noninteracting fermions through a periodically driven quantum point contact~\cite{Gamayun21}. Recently a nanomechanical system consisting of three harmonic lattices---each one acted upon by a pinning harmonic potential---where the central segment is coupled to the other two through a time-dependent mechanical interaction similar to that herein employed has been studied by means of the Keldysh nonequilibrium Green function formalism~\cite{Beraha22}. The system presents different stationary transport regimes depending on the values of the driving frequency and of the imposed static thermal bias, as well as on the degree of locality of the interactions. It would be interesting to explore the possibility of obtaining the same effects with the anharmonic model herein employed. Another extremely interesting direction could be to extend the system in two dimensions, which could provide new insights as in the case of thermal rectification~\cite{Lan06}. Finally we mention that recently it was reported an implementation of the FK model with cold Rydberg dressed atoms in an optical lattice~\cite{Munoz20}. By using the Rydberg dressing and realistic experimental parameters it is possible to realize the springlike interaction potential, similar to the one employed in original FK model and in this work. This proposal could offer an interesting possibility to implement the herein periodic external driving in a more realistic system if the problem of the short lifetimes of the Rydberg states could be overcomed. 

\smallskip
\begin{acknowledgments}
The author thanks Consejo Nacional de Humanidades Ciencias y Tecnolog\'\i as (CONAHCYT) Mexico for financial support, B.~A.~Mart\'\i nez-Torres for his help in processing the data, and Juan~M. L\'opez for insightful comments and discussions, as well as for allowing access to the computer facilities of the Instituto de F\'\i sica de Cantabria (IFCA), Spain where most simulations were performed.
\end{acknowledgments}

%%%%%%%%%%%%%%%%%%%%%%%%%%%%%%%%%%%%%%%%%%%%%%%%%%%%%%%%%%%%%%%%%%%%%%%%%%%%%%%%%%%%%%%%%%%%%%%%%%%%%%%%%%%%%

\bibliographystyle{prsty}
%\bibliography{Bibliography}

\begin{thebibliography}{10}

\bibitem{Yan11}
H.~Yan, H.~S.~Choe, S.~Nam, Y.~Hu, S.~Das, J.~F.~Klemic, J.~C.~Ellenbogen, and C.~M.~Lieber, Nature {\bf 470},  240  (2011).

\bibitem{Lu07}
W.~Lu and C.~M.~Lieber, Nature Mater. {\bf 6},  841  (2007).

\bibitem{Yan22}
Q.~Yan and M.~G.~Kanatzidis, Nature {\bf 21},  503  (2022).

\bibitem{Cuevas10}
J.~C.~Cuevas and E.~Scheer, {\em Molecular Electronics} (World Scientific,
  Singapore, 2010).

\bibitem{Xuefeng14}
G.~Xuefeng, AIMS Materials Science {\bf 1},  11  (2014).

\bibitem{Su16}
T.~A.~Su, M.~Neupane, M.~L.~Steigerwald, L.~Venkataraman, and C.~Nuckolls, Nat. Rev. Mater. {\bf 1},  16002  (2016).

\bibitem{Segal03}
D.~Segal, A.~Nitzan, and P.~H\"anggi, J. Chem. Phys. {\bf 119},  6840  (2003).

\bibitem{Braun04}
O.~M. Braun and Y.~S. Kivshar, {\em The Frenkel-Kontorova model: Concepts, methods, and applications} (Springer-Verlag, Berlin Heidelberg, 2004).

\bibitem{Bao-quan10}
B.~Q. Ai, D.~He, and B.~Hu, Phys. Rev. E {\bf 81},  031124  (2010).

\bibitem{Cuansing10}
E.~C.~Cuansing and J.-S.~Wang, Phys. Rev. E {\bf 82},  021116  (2010).

\bibitem{Zhang11}
S.~Zhang, J.~Ren, and B.~Li, Phys. Rev. E {\bf 84},  031122  (2011).

\bibitem{Romero20}
M.~Romero-Bastida, Phys. Rev. E {\bf 102},  052124  (2020).

\bibitem{Guo11}
Z.-X.~Guo, D.~Zhang, and X.-G.~Gong, Phys. Rev. B {\bf 84},  075470  (2011).

\bibitem{Laxmi23}
V.~Laxmi, N.~Basu, and P.~K.~Nayak, J. Raman Spectrosc. {\bf 54},  76  (2023).

\bibitem{Ong11}
Z.-Y.~Ong and E.~Pop, Phys. Rev. B {\bf 84},  075471  (2011).

\bibitem{Zhu23}
X.~Zhu {\it et~al.}, J. Phys. Chem. C {\bf 127},  3246  (2023).

\bibitem{Varshney12}
V.~Varshney {\it et~al.}, Nanoscale {\bf 4},  5009  (2012).

\bibitem{Sun13}
T.~Sun, J.~Wang, and W.~Kang, Nanoscale {\bf 5},  128  (2013).

\bibitem{Li12}
N.~Li, J.~Ren, L.~Wang, G.~Zhang, P.~H\"anggi and B.~Li, Rev. Mod. Phys. {\bf 84},  1045  (2012).

\bibitem{Li04a}
B.~Li, L.~Wang, and G.~Casati, Phys. Rev. Lett. {\bf 93},  184301  (2004).

\bibitem{Li05}
B.~Li, J.~H.~Lan, and L.~Wang, Phys. Rev. Lett. {\bf 95},  104302  (2005).

\bibitem{Zhang17}
L.~Zhang and L.~Liu, ACS Applied Materials \& Interfaces {\bf 9},  28949
  (2017).

\bibitem{Dong19}
Y.~Dong, C.~Diao, Y.~Song, H.~Chi, D.~J.~Singh, and J.~Lin, Phys. Rev. Appl. {\bf 11},  024043  (2019).

\bibitem{Li05a}
B.~Li, J.~Wang, L.~Wang, and G.~Zhang, Chaos {\bf 15},  015121  (2005).

\bibitem{Zhang18}
X.~Zhang, K.~Ma, J.~Zhang, and L.~Nie, Eur. Phys. J. B {\bf 91},  317  (2018).

\bibitem{Luo07}
C.~Luo and J.~Dong, Phys. Rev. B {\bf 75},  132301  (2007).

\bibitem{Gamayun21}
O.~Gamayun, A.~Slobodeniuk, J.-S.~Caux, and O.~Lychkovskiy, Phys. Rev. B {\bf
  103},  L041405  (2021).

\bibitem{Beraha22}
N.~Beraha, A.~Soba, and M.~F.~Carusela, J. Phys. A: Math. Theor. {\bf 55},
  225304  (2022).

\bibitem{Lan06}
J.~Lan and B.~Li, Phys. Rev. B {\bf 74},  214305  (2006).

\bibitem{Munoz20}
J.~M. Mu\~noz, R.~Sawant, A.~Maffei, X.~Wang, and G.~Barontini, Phys. Rev. A {\bf 102},  043308  (2020).

\end{thebibliography}

\end{document}